\documentclass[runningheads]{llncs}
\usepackage{graphicx}
\usepackage{listings}
\usepackage{stmaryrd}
\usepackage{xcolor}
\usepackage{graphicx}
\usepackage{float}
\usepackage{tikz}
\usepackage{booktabs}
\usepackage{url}
\usepackage{tabularx}
\usepackage{array}
\usepackage{parskip}

\definecolor{isarblue}{HTML}{006699}
\definecolor{isarlight}{HTML}{0099ff}
\definecolor{isargreen}{HTML}{009966}
\definecolor{isarred}{HTML}{e34013}
\definecolor{lightgray}{gray}{0.95}
\lstdefinelanguage{isabelle}{%
    keywords=[1]{type_synonym,datatype,primrec,fun,abbreviation,definition,proof,lemma,theorem,corollary,value,using,by,proposition, record, locale, inductive_set, unfolding, from, moreover,ultimately,consider, qed, next, then, have},
    keywordstyle=[1]\bfseries\color{isarblue},
    keywords=[2]{fixes,where,assumes,shows,and,obtains,for},
    keywordstyle=[2]\bfseries\color{isargreen},
    keywords=[3]{fix, assume, case, define, show},
    keywordstyle=[3]\color{isarlight},
    commentstyle=\color{isarred},
    morecomment=[s]{(*}{*)},
    moredelim=[is][\color{isarblue}]{|-}{-|}, 
    alsoletter={?},
}
\newcommand*{\circled}[2][]{\tikz[baseline=(C.base)]{
    \node[inner sep=0pt] (C) {\vphantom{1g}#2};
    \node[draw, circle, inner sep=3pt, yshift=1pt]
        at (C.center) {\vphantom{1g}};}}

\begin{document}
\title{Towards Automating Blockchain Consensus Verification with IsabeLLM}
%
%
\author{Elliot Jones\inst{1} \and William Knottenbelt}
\authorrunning{E. Jones and W. Knottenbelt}
%
\institute{Department of Computing, Imperial College London, United Kingdom\\
\inst{1}\email{e.jones24@imperial.ac.uk}} 
\maketitle              
\begin{abstract}
Consensus protocols are crucial for a blockchain system as they are what allow agreement between the system's nodes in a potentially adversarial environment. For this reason, it is paramount to ensure their correct design and implementation to prevent such adversaries from carrying out malicious behaviour. Formal verification allows us to ensure the correctness of such protocols, but requires high levels of effort and expertise to carry out and thus is often omitted in the development process. In this paper, we present IsabeLLM, a tool that integrates the proof assistant Isabelle with a Large Language Model to assist and automate proofs. We demonstrate the effectiveness of IsabeLLM by using it to develop a novel model of Bitcoin's Proof of Work consensus protocol and verify its correctness. We use the DeepSeek R1 API for this demonstration and found that we were able to generate correct proofs for each of the non-trivial lemmas present in the verification.

\end{abstract}
\section{Introduction}

A blockchain enables peer-to-peer digital transactions without the need for a trusted intermediary. This is only possible because of its consensus protocol, which allows nodes within the system to agree on the state of the blockchain, even in the presence of adversaries. For this reason, it is paramount that consensus is designed and implemented correctly to prevent the system from reaching unwanted states that can be exploited by adversaries. The most famous example of this is Bitcoin's Proof of Work (PoW) consensus protocol and its susceptibility to a 51\% attack, where adversaries control the majority of the compute power in the system, which gives them the potential to double spend. Infamous examples of such attacks include Ethereum Classic~\cite{EthClassic51}, Bitcoin Gold~\cite{BTCgold51}, and Vertcoin~\cite{VertCoin51}, totalling losses of over \$30 million. 

Other key components of modern blockchain systems are bridging protocols for cross-chain data transfer and smart contracts for automated agreement execution. These components are also not without their exploits, with infamous examples such as the Poly Network~\cite{PolyNetworkAttack}, Wormhole Bridge~\cite{WormholeAttack}, Binance Smart Chain~\cite{BNBChainAttack}, and Qubit Finance~\cite{QubitFinAttack}, totalling losses of over \$1.5 Billion. These failures further underscore the need to ensure correctness across the domain.

Formal verification is the process of formalising a system and then mathematically proving its correctness. However, it is often underutilised in the software development process because of the large amount of effort and expertise it requires. The blockchain domain is no exception to this, resulting in the huge financial losses discussed previously. Furthermore, blockchain systems cannot rely on the traditional \lq test and patch\rq model for their consensus protocols as patching would require a hard fork, such as the Ethereum/Ethereum Classic split~\cite{EthHardFork}. Hard Forks are extremely disruptive and controversial as they challenge blockchain's core principle of immutability. Furthermore, patching smart contracts is often impossible once they have been deployed on a blockchain as they are usually immutable~\cite{SmartContractsImmutable} with the exception of some upgradeable contracts~\cite{SmartContractsUpgradeable}. This further amplifies the need for formal verification, as it can be used for correctness-by-construction~\cite{CorrectByConstruct} and minimise the need for costly post-deployment fixes.

In recent years, the field of Artificial Intelligence has made incredible progress, particularly within the realm of Large Language Models (LLMs) like OpenAI's ChatGPT and High-Flyer's DeepSeek. This advancement has opened up new opportunities across all domains, including the formal verification space. In particular, AI for theorem proving has gained traction and has started to see applications outside of purely mathematical statements and instead for program verification. An example of this is FVEL~\cite{aitpFVEL}, which is used to assist in automated verification of C/C++ programs in the Isabelle proof assistant. This reduces the entry barrier for the formal verification of such programs, making it more accessible and less time-consuming.

In this paper, we present IsabeLLM, a tool that integrates the proof assistant Isabelle with a Large Language Model to assist and automate proofs. We demonstrate the effectiveness of IsabeLLM by using it to prove the correctness of a novel model for Bitcoin's Proof-of-Work consensus protocol. The contributions of this paper are as follows:

\begin{enumerate}
    \item The IsabeLLM tool, which can be used with any LLM API and is general purpose, allowing it to be used for theorem proving within any domain. In this paper, we focus on verifying blockchain consensus. We describe IsabeLLM's architecture (Section~\ref{sec:arch}) and implementation (Section~\ref{sec:imp}).
    \item A novel mechanised model of Bitcoin's Proof of Work consensus protocol in Isabelle, with correctness proven using IsabeLLM. The model is an extension of the work done in~\cite{ElliotDiego}, which we describe in Section~\ref{sec:mod}. We use DeepSeek R1 as our chosen LLM to integrate with IsabeLLM.
    \item Analysis of the performance of IsabeLLM, looking at success rate, number of iterations, and any emerging pain points. We describe the results in Section~\ref{sec:res}.
\end{enumerate}

\section{Background}

\subsection{Blockchain}

A blockchain is a decentralised ledger that allows two parties to carry out transactions without the need of a trusted intermediary, eliminating the need for trust. This is only possible through a blockchain's consensus protocol, which allows all parties to agree on the current state of the blockchain and the transactions recorded on it. The most popular consensus protocol is Proof of Work (PoW) used by Bitcoin's blockchain, which has around 1.2 billion recorded transactions~\cite{BtcTransactions} with Bitcoin's market capitalisation sitting around \$1.8 trillion~\cite{BtcMarketCap}. The core idea of PoW is that the longest blockchain is correct since it assumes the majority of computing power within the system is honest and therefore should be able to solve hashes and add blocks faster than adversaries~\cite{BtcWhitepaper}.

\subsection{Isabelle}

Isabelle is a proof assistant written in Scala and ML that uses Higher-Order Logic (HOL). It is used to write and verify formal proofs with high assurance due to the mechanisation of these proofs~\cite{Isabelle}. Isabelle's Isar proof language allows these proofs to be more readable than the traditional approach to theorem proving by repeatedly applying tactics. Isabelle also makes use of automation tools like Sledgehammer, which uses external automated theorem provers (ATPs) to help you complete proofs. Outside of the proof assistant itself, the Scala library Scala-Isabelle provides the functionality to interact with an Isabelle process inside of a Scala application~\cite{scalaIsabelle}.

\section{Related Work}

Isabelle has been used extensively in the last 20 years to carry out numerous verifications. Some of the most notable verifications include the seL4 Microkernel~\cite{isabelleSel4}, the ML compiler~\cite{isabelleCakeML}, and numerous protocol and program verifications~\cite{isabelleDistributedSystems,isabelleHybridSystems,isabelleSAT}. Outside of verification, Isabelle has been used to formalise a large amount of mathematics that can be found in the Archive of Formal Proofs (AFP)~\cite{isabelleAFP}. Some of the most notable formalisms in the AFP include G{\"o}del's incompleteness theorems~\cite{isabelleGodel}, Jordan curve theorem~\cite{isabelleJordanCurve}, and Ramsey's Theorem~\cite{isabelleRamsey}. In recent years, Isabelle has been used for verifications and formalisms of blockchain systems, including the Ethereum Virtual Machine~\cite{isabelleEVM} and a framework to verify solidity smart contracts~\cite{isabelleSolidity}.

Outside of Isabelle, various other theorem provers have been used to carry out verifications in the blockchain domain. To name a few, Agda~\cite{agdaBtc,agdaBtcScript,agdaSolidity}, Coq~\cite{coqSmartContract,coqEthSmartContract,coqAlgorand}, and Lean~\cite{leanClear,LeanAMM} have been used for the formalisms of blockchain. The field has also started to see formalisms of Decentralised Finance (DeFi) specific components~\cite{FormalDeFi,FormalAMM,FormalEV} but are yet to be mechanised. Other major works within the space include KEVM~\cite{kevm}, Certora Prover~\cite{certoraProver}, and Mythril~\cite{mythril}.

The field of AI for theorem proving has seen the development of major data sets for proof assistants in recent years, including IsarStep and PISA for Isabelle~\cite{aitpIsarStep,aitpPISA}, LeanDojo for Lean~\cite{aitpLeanDojo}, and GamePad and CoqGym for Coq~\cite{aitpCoqGym}. Using these datasets has allowed for the development of various theorem proving models, including LEGO-Prover~\cite{aitpLegoProver}, LISA~\cite{aitpPISA} and DeepSeek-Prover~\cite{aitpDeepSeekProver}. Artificial Intelligence for formal verification has seen limited use, with the aforementioned FVEL~\cite{aitpFVEL} being the major work in this area. As for AI for formal verification of blockchain, the literature is sparse and has only seen research into extracting smart contract specifications from natural language~\cite{aitpSmartContractSpecs} to the best of our knowledge.

\section{Model}
\label{sec:mod}

\begin{table}[htbp]
\centering
\begin{tabular}{@{} l l@{\hskip 20pt}l @{}}
\toprule
\textbf{Lemma Name} & \textbf{Binary Tree} & \textbf{N-ary Tree} \\
\midrule
subtree\_height & N/A & 15 \\
height\_mono & 1+1 & 23 \\
obtain\_max & N/A & 23 \\
foldr\_max\_eq & N/A & 37 \\
branch\_height & N/A & 30 \\
sub\_longest & N/A & 28 \\
sub\_branch & N/A & 41 \\
weaken\_distance & 1 & 18 \\
weaken\_depth & 1 & 15 \\
common\_prefix & 25+12 & 38 \\
height\_add (mining) & 10+5 & 36 \\
check\_add (mining) & 49+158 & 1 \\
height\_add (honest) & 10+5 & 32 \\
check\_add (honest) & 22+13 & 36 \\
bounded\_check & 56 & 17 \\
consensus & 1 & 5 \\
\textbf{Total} & \textbf{175+193} & \textbf{395} \\
\bottomrule
\end{tabular}
\vspace{10pt}
\caption{Lines of Proof (LoP) for each tree model.}
\label{tb1}
\end{table}

Our consensus model builds on previous work~\cite{ElliotDiego} by generalizing the blockchain structure from a binary tree to an n-ary tree. This extension enables the model to account for an arbitrary number of forks in any given block, reflecting a more realistic view of a blockchain. We prove that consensus holds in a majority honest network using the common prefix and chain quality properties outlined in the Bitcoin Backbone Protocol~\cite{btcBackbone}, where they are discussed in more detail. We make the same assumptions of majority honesty and synchronisation in the network, meaning that the majority of the computing power in the network is honest and that everyone shares the same view of the blockchain. As in the previous work, we can omit the chain quality property under our majority honesty assumption, leaving us with the common prefix property which states that all honest parties agree on a common chain up to the last {$k$} blocks in a chain. This is a safety property, showing honest nodes do not diverge except near the tip of the chain. Its implementation in our Isabelle model can be seen in Fig.~\ref{isabellefig}.

\begin{figure}[htbp]
\centering
\begin{minipage}{\linewidth}
\begin{lstlisting}
theorem consensus:
  fixes t assumes "t \<in> traces"
  and "p \<in> longest (State (hd t))"
  and "p' \<in> longest (State (hd t))"
shows "take k p = take k p'"
\end{lstlisting}
\end{minipage}
\caption{Consensus theorem in Isabelle}
\label{isabellefig}
\end{figure}

This statement is identical to the consensus statement for the binary tree model. However, the generalisation to an n-ary tree significantly increases the complexity of the proof. In the binary tree case, inductive arguments typically require only two cases (e.g., left and right subtrees), whereas the n-ary setting necessitates reasoning over an arbitrary number of branches, complicating case distinctions and inductive reasoning. To show this, Table~\ref{tb1} shows the Lines of Proof (LoP) required to complete the verification of each model. We only list the lemmas that were more than one LoP in at least one of the models. In the binary tree column, ``N/A'' means that the lemma was not required for the verification. For the rows with \(x+y\), \(x\) is the LoP that are `original' and \(y\) is the LoP that are symmetric to $x$ and are just repeated for the different cases. With this in mind, it is clear that the n-ary tree model has more than double the original LoP when compared to the binary tree model.    

\section{IsabeLLM}

\begin{figure}[htbp]
  \centering
  \includegraphics[width=0.8\linewidth]{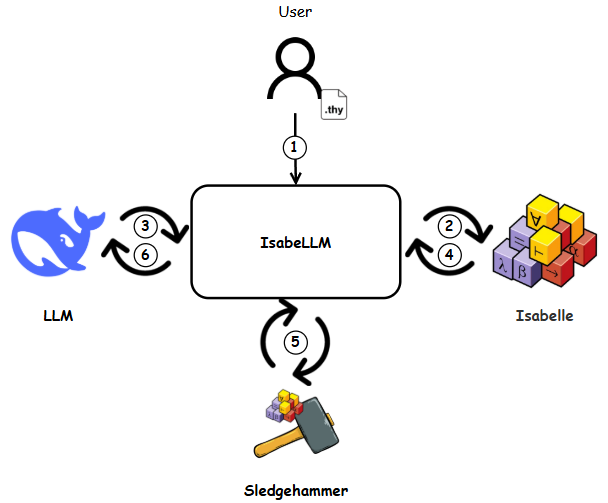}
  \caption{IsabeLLM Architecture.}
  \label{fig:arch}
\end{figure}

IsabeLLM is an interface between the Isabelle proof assistant and an LLM. It is designed for general purpose and so can be used to prove any kind of statements within Isabelle. It should be noted that if you are using bespoke imports for your theory file, then they should be given to the LLM as context for it to understand. In our models, we are only importing Isabelle's Main library, and everything is contained within the single theory file, meaning we do not need to provide extra context.

\subsection{Architecture}
\label{sec:arch}
 
The high-level architecture for IsabeLLM can be seen in Fig.~\ref{fig:arch}. The main idea is that we use an LLM to understand the high-level structure of a proof and then use Isabelle's Sledgehammer tool to solve the intermediate steps that the LLM failed (if any). The general workflow for IsabeLLM is as follows:

\begin{enumerate}
    \item [\circled{1}] The user uploads their Isabelle theory file (.thy) to their working directory, along with a ROOT file so that the Isabelle server knows which files to look at. The user starts IsabeLLM.
    \item[\circled{2}] IsabeLLM first uses the Isabelle server to try and build the theory file. If there are no issues with the file and all statements have been proven, then the build completes, and we are done. If not, then IsabeLLM captures the errors raised to identify the unproven statements and extracts them.
    \item[\circled{3}] IsabeLLM sends the context of the theory file and the unproven lemma to the LLM via its API. The LLM tries to prove the lemma and returns a proof of the statement.
    \item[\circled{4}] IsabeLLM injects the new proof into the theory file and tries to build it again. If this fails, we send the file to Isabelle's Sledgehammer tool.
    \item[\circled{5}] Sledgehammer tries to solve each unproven line within the proof. If some are left unproven, then IsabeLLM extracts these lines and their errors, along with the rest of the updated theory file.
    \item[\circled{6}] IsabeLLM returns the current proof state to the LLM and asks it to resolve the remaining errors. The LLM returns a proof of the statement.
    \item [\circled{7}] Steps 4--6 are repeated until the theory file is successfully built or IsabeLLM reaches a set number of iterations.
\end{enumerate}

\begin{figure}[htbp]
  \centering
  \includegraphics[width=0.7\linewidth]{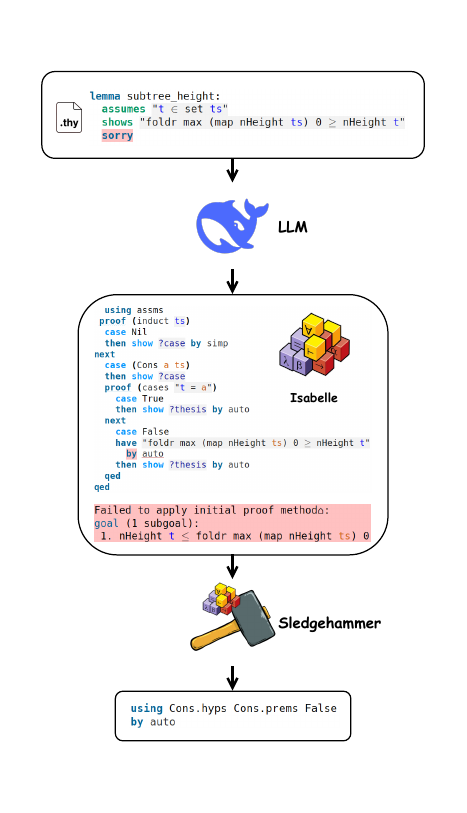}
  \caption{IsabeLLM example workflow.}
  \label{fig:example}
\end{figure}

Fig.~\ref{fig:example} shows an example workflow in IsabeLLM. In this example, we prove the lemma subtree\_height which states that the height of a tree in a set of trees is always greater than or equal to the maximum height of the set of trees. The LLM generates a proof that fails a proof step, which we then correct with Sledgehammer. The proof is by induction over the list of trees and splitting the inductive step into the cases of whether the tree is at the head of the list or not.

For this paper, we opted to use DeepSeek R1 as our LLM due to its strong coding benchmarks and free access to its API. Claude Sonnet was also considered, but was ultimately decided to be too expensive. Other models like OpenAI's GPT-4 and Mistral's Le Chat were also considered but showed poor performance during manual testing. As for our choice of proof assistant, we chose Isabelle due to its existing automation tool Sledgehammer and integration library Scala-Isabelle. Although not integral, the Isar language also helps to understand the logic of the proofs and the dialogue between IsabeLLM and DeepSeek. We also note that we use Isabelle2022 as it is compatible with Scala-Isabelle.

\begin{table}[htbp]
\centering
\begin{tabularx}{\textwidth}{|>{\hspace{6pt}}l<{\hspace{6pt}}|>{\hspace{6pt}}X<{\hspace{6pt}}|}
\hline
\textbf{Error} & \textbf{Description} \\
\hline
``Sorry'' Detected in Proof & 
The sorry keyword is used to mark incomplete proofs. IsabeLLM uses sorry to detect on which part of the theory file to call the LLM.
\\
\hline
Failed Proof &
The proof method/tactic failed to complete the goal. This means the generated proof was incorrect. We first attempt to Sledgehammer the proof before returning to the LLM.
\\
\hline
Undefined Fact/Method &
Usually means the LLM has hallucinated a fact, method, or attribute of either that does not exist. We simply remove these hallucinations and try to rebuild.
\\
\hline
Lexical/Syntax Error &
Bad syntax injected into the theory file. IsabeLLM has various methods for trying to detect these issues and resolving them. If IsabeLLM fails to resolve the issue, it will either return the proof back to the LLM or cancel the computation and ask the user to make amends. The latter is usually reserved for when the LLM gives a very incoherent answer, such as DeepSeek responding in Mandarin or looping, which is uncommon.
\\
\hline
Timeout &
The theory file does not build in the allotted time. This usually means there is a hanging proof step that is not evaluating correctly. The metis and blast tactics are common culprits for this. IsabeLLM searches through the modified proofs for occurrences of these tactics and calls Sledgehammer on each of them to check if they are being evaluated correctly.
\\
\hline
\end{tabularx}
\vspace{10pt}
\caption{Isabelle build errors.}
\label{tb2}
\end{table}

\subsection{Implementation}
\label{sec:imp}

Almost all of IsabeLLM is written in Scala, with some Python to access the LLM API using the openai library. The main reason for choosing Scala is to be able to use the Scala-Isabelle library, which offers the functionality to control an Isabelle process from a Scala application. In particular, we make use of Scala-Isabelle for calling Sledgehammer in our theory file. Our approach to using Scala-Isabelle was inspired by the work done on PISA~\cite{aitpPISA}. All of our code, including the IsabeLLM source code and theory files, can be found at~\cite{isabellmRepo}. We list IsabeLLM's features below:

\begin{enumerate}
    \item Interface between Isabelle and a LLM API.
    \item Code extraction from a theory file, including lemmas, definitions, and proofs.
    \item Injection of code into a theory file, including lemmas, definitions, and proofs.
    \item Sledgehammer functionality with a timeout control and option to select which provers it uses. In this paper, we set this timeout to 60 seconds and use the default provers.
    \item Handling of errors in the build process.
    \item Records and updates to the LLM chat history.
    \item LLM Prompt Generation.
    \item Workflow for automated theorem proving using all of the above. The workflow has a control for the maximum number of LLM iterations before timing out to prevent endless loops. 
    
\end{enumerate}

Due to the stochastic nature of LLMs, the most challenging part of automating proof with IsabeLLM is ensuring the output from the LLM has the correct syntax to be injected into Isabelle. Generally speaking, most models understand the syntax well enough to give a coherent response. However, most outputs will trigger at least one error in the build process and must be handled accordingly. Table~\ref{tb2} highlights the general types of error that are encountered when trying to build the theory file after injecting a generated proof.

When sending requests to the LLM, IsabeLLM automatically builds the required prompts to make the context clearer. When we first initialise a proof, we send a prompt that includes the context of the lemma we are trying to prove and everything in the theory file before it. After the initialisation prompt, we send prompts that include only the current proof state of the lemma. The templates for these prompts can be seen in Table~\ref{prompts}. IsabeLLM also maintains the chat history with the LLM. To minimise the size of our context, we reset the history after successfully proving a lemma, then update the initial context to the theory file with the updated lemma. A JSON file containing the chat history for each lemma is stored.

\begin{table}[htbp]
\centering
\begin{tabularx}{\textwidth}{|>{\hspace{6pt}}l<{\hspace{6pt}}|>{\hspace{6pt}}X<{\hspace{6pt}}|}
\hline
\textbf{Prompt} & \textbf{Text} \\
\hline
Initialisation & 
I am trying to complete a proof in Isabelle. Here is my theory file so far:  (.thy file). I am trying to prove the following lemma: (lemma). Please prove this lemma. Return only the raw code without any additional text, explanations, formatting, or commentary. Do not include ``` or language tags. Just the pure code.
\\
\hline
Error &
Your proof is incorrect. The current proof state is: (proof state). The line: (error line) produced the following error message: (error). Please amend the proof to deal with this error. Return only the raw code without any additional text, explanations, formatting, or commentary. Do not include ``` or language tags. Just the pure code.
\\
\hline

\end{tabularx}
\vspace{10pt}
\caption{IsabeLLM prompts.}
\label{prompts}
\end{table}

\section{Results}
\label{sec:res}

To test the effectiveness of IsabeLLM, we try to prove each of the 16 lemmas listed in Table~\ref{tb1} 10 times with a maximum of 5 iterations per attempt, not counting instances when the LLM would return an empty response. It should be noted that the LoP specified for each lemma can vary as the LLM can generate different proofs for the same thing. As mentioned previously, we used the DeepSeek R1 API for this experiment. We consider an attempt to be a failure if it exceeds the maximum number of iterations or exits prematurely (often due to syntax issues). Table~\ref{tb3} shows the results of using IsabeLLM for each lemma. 

Generally speaking, IsabeLLM was able to prove each lemma multiple times, often with a varying number of iterations required to do so. Some lemmas, like subtree\_height, were repeatedly solved with one iteration but almost always required intervention to amend either incorrect syntax or Sledgehammer incorrect proof steps. The general approach of the proofs generated for these lemmas was always very similar and showed little variation. This probably stems from the fact that these lemmas usually had shorter proofs on average, leaving less room for variety.

As expected, we tended to see fewer successful attempts for the larger proofs, which left more room for variety. The lemmas with which it seemed to struggle most was branch\_height and bounded\_check. We found that Sledgehammer struggled to resolve the intermediate steps provided by the LLM despite the fact that the high-level was sound. Interestingly, we repeated Sledgehammer on these steps in Isabelle2025 and they were solved every time, highlighting IsabeLLM's potential to evolve with Isabelle.

As for the failed attempts, we saw a recurring pattern in which the LLM would fixate on a proof step and disregard the proof as a whole. In particular, when a proof step would fail and IsabeLLM was unable to find a proof with Sledgehammer, we would send this error back to LLM. The LLM would often just repeat the same proof back to us with a slightly modified proof of the step, which would very rarely succeed if Sledgehammer had already failed to find one. This would create a loop of tweaking and failing the same step without progress. IsabeLLM was most successful when the LLM broke the proof step down into more manageable parts, which Sledgehammer could then solve itself.

\begin{table}[htbp]
\centering
\begin{tabularx}{\textwidth}{@{} l >{\centering\arraybackslash}X >{\centering\arraybackslash}X >{\centering\arraybackslash}X @{}}
\toprule
\textbf{Lemma Name} & \textbf{Successful Attempts} & \textbf{Avg. Iterations (Success)} & \textbf{Lines of Proof} \\
\midrule
subtree\_height         & 10 & 1   & 15 \\
height\_mono            & 10 & 1   & 23 \\
obtain\_max             & 9  & 1.4 & 23 \\
foldr\_max\_eq          & 5  & 2   & 37 \\
branch\_height          & 3* & 2   & 30 \\
sub\_longest            & 7  & 1.1 & 28 \\
sub\_branch             & 5  & 1.8 & 41 \\
weaken\_distance        & 10 & 1   & 18 \\
weaken\_depth           & 10 & 1   & 15 \\
common\_prefix          & 6  & 1.5 & 38 \\
height\_add (mining)    & 6  & 2   & 36 \\
check\_add (mining)     & 10 & 1   & 1 \\
height\_add (honest)    & 8  & 1.7 & 32 \\
check\_add (honest)     & 9  & 1.2 & 36 \\
bounded\_check          & 7* & 1   & 17 \\
consensus              & 10 & 1   & 5 \\
\bottomrule
\end{tabularx}
\vspace{10pt}
\caption{Number of successful proof attempts}
\label{tb3}
\begin{minipage}{\textwidth}
\vspace{3pt}
\small\textit{* Indicates these proofs were completed using Sledgehammer from Isabelle2025.}
\end{minipage}
\end{table}

\section{Discussion}

One limitation we faced was the speed and occasionally unreliable API for DeepSeek R1. We used the free OpenRouter API for this work and found it to be considerably slow at times. This was to be expected with the free API, as OpenRouter also offers a paid version with improved latency and tokens per second. An unexpected issue was that the API would occasionally return an empty output, forcing us to add a condition to handle this and repeat the iteration. We expect these limitations to be mitigated with an improved API or by running the LLM locally. Another expected issue was hallucinations of the LLM. It is common for the LLM to get Isabelle's syntax wrong, use a non-existent theorem, or try to prove something that was impossible. Many of these issues were handled in the workflow, as discussed in Section~\ref{sec:imp}, but sometimes manual intervention was required to sort out the issues before resuming the computation. Unfortunately, there is not much that can be done here, but we expect this issue to minimise with time as LLMs and Sledgehammer improve.

A key area of improvement would be to improve the efficiency of Sledgehammer. As mentioned previously, IsabeLLM runs on Isabelle2022 and so does not benefit from the improved Sledgehammer in later releases. This highlights the need to make IsabeLLM compatible with later Isabelle releases. We also found that calling Sledgehammer remotely for IsabeLLM does not generate counterexamples for impossible proofs. These are usually detected by nitpick, an internal Sledgehammer tool, before running the provers on the step. This would save us from wasting computation time on impossible proofs and also allow us to give more context to the LLM. Furthermore, we found that Sledgehammer would be repeatedly called on the same proof steps between iterations as the LLM would ignore our new proof step and go back to using the incorrect step it gave us from a previous iteration. With this in mind, it would be effective to incorporate functionality that detects repeated steps and stores the correct proof so that it can be injected quickly without having to run again. The final issue with Sledgehammer was its high memory use, particularly when there were back-to-back calls, usually when consecutive generated proof steps are incorrect. We would find that the first call would still use significant memory when the next Sledgehammer was called, causing our machine to run out of memory and killing the process. This issue is largely internal for Sledgehammer and is out of IsabeLLM's control, but again this should improve with later Isabelle releases.

IsabeLLM's main limitation as a proof automation tool is that it only automates the proof of statements, not the generation of the statements themselves. This means that the user must specify a statement before it can be proven, including other key parts of the theory, such as functions, locales, and sets. However, this issue will be largely mitigated if IsabeLLM is used in conjunction with existing blockchain verification frameworks, rather than building from the ground up like our model. For example, Isabelle/Solidity~\cite{isabelleSolidity} builds most of the model automatically, and you only have to specify the invariant property you are trying to verify for a given smart contract. A challenge that comes with this is for the LLM to understand the bespoke calculus that comes with such frameworks, as there will be far fewer proof corpora to learn from. 

\section{Conclusion}

In this paper, we introduce the proof automation tool IsabeLLM for Isabelle proof assistant. We then used IsabeLLM to complete a novel verification of PoW consensus and analysed its effectiveness.

An area of future work would be to modify IsabeLLM so that it constructs a proof tree by querying the LLM in parallel and branching the proof in different directions for each different proof the LLM gives. This is the standard method used in the field for AI for theorem proving~\cite{aitpDeepSeekProver,aitpPISA} and would help prevent IsabeLLM from getting stuck in a loop of repeatedly trying to prove the same step. This could be taken further by using different LLMs, which would likely generate different approaches to the proof. As the field progresses, more advanced models like Claude Opus 4 are likely to replace our choice of DeepSeek.

Another area of work is using IsabeLLM for more complex proofs that are not necessarily within the blockchain domain and split across multiple theory files. As mentioned previously, IsabeLLM is designed for general purpose and so can be used for proofs in any domain. Furthermore, LLMs could also be fine-tuned on proof corpora datasets like the Archive of Formal Proofs to see how it improves performance. Alternatively, instead of an LLM, bespoke language models could be created and used for Isabelle, such as the work done on LISA~\cite{aitpPISA} that used AFP as a training set.

Lastly, auxiliary techniques like Retrieval-Augmented Generation (RAG) or Static Prompt Templating could be employed to mitigate our issue of the LLM re-attempting failed proof steps with minimal variation and hallucinations as a whole. Doing so could make IsabeLLM more robust and less dependent on the quality of the LLM itself.

IsabeLLM shows great promise towards automated verification and will only improve in ability as LLMs and Sledgehammer continue to evolve. Further work on IsabeLLM's functionality to handle different syntax errors from generated proofs could also help to improve the speed and reliability of the automation process.

\appendix
\section*{Appendix}

All relevant code for IsabeLLM can be found at:

\begin{center}
\url{https://github.com/EllbellCode/IsabeLLM}
\end{center}

The repository includes:
\begin{itemize}
  \item Source code for IsabeLLM.
  \item Isabelle theory files for the n-ary tree PoW model.
  \item Setup instructions.
\end{itemize}

\clearpage
\bibliographystyle{splncs04}
\bibliography{Main}

\end{document}